\documentclass[aps,twocolumn,showpacs]{revtex4}

\begin{document}

\title{Can dark matter be mostly massless particles?} 

\author{Xiang-Song Chen} 
\affiliation{Department of Physics, Sichuan University,
                Chengdu 610064, China; and\\
	Joint Center for Particle, Nuclear Physics, and Cosmology, 
                Nanjing University, Nanjing 210093, China\\
 {\rm Electronic address: cxs@scu.edu.cn, cxs@chenwang.nju.edu.cn}}
\date{October, 2005}
                                                                            
\begin{abstract}
We discuss an overlooked factor in dark matter studies. Namely, if massless particles are captured into a local structure and stop free streaming in the universe, they no longer lose energy by cosmological red-shift, and no longer smear out density fluctuations beyond their ``confinement'' scale. If this occurred at the stage when radiation dominated over baryonic matter in energy density, then these captured massless particles would comprise the major part of dark matter in today's universe, leaving no room for other dark matter scenarios. The most probable such particles are gravitons with non-linear self-interaction. 
\end{abstract}
\pacs{95.35.+d}
\maketitle

The nature of dark matter (DM) in galaxies and clusters of galaxies is one of the most puzzling problems in cosmology. \cite{Sahni} The limit on baryon density inferred from nucleonsythesis tells that DM must be mostly non-baryonic. When some kind of particle is considered as the major component of DM, a significant mass is always required for this particle. 
The reason seems self-evident: As massless particles travel through the expanding universe, their energy drops by red-shift, and becomes negligible in today's matter-dominated era. It is often required further that the mass of DM particles should be large enough to make them ``cold'' (non-relativistic), because the free streaming of ``hot'' (relativistic) particles would smear out density fluctuations at small scales, and makes galaxy formation too slow.

We point out, however, a primitive but vital factor which apparently has been universally overlooked. Namely, massless particles do not always stream freely in the universe, they can also get captured into a local structure. 
Once this happens, they no longer lose energy through cosmological red-shift, and no longer smear out density fluctuations beyond their ``confinement'' scale. One may readily realize that this would give massless particles an opportunity to be the major part of DM today: The early universe was dominated by radiation, if a significant fraction of massless particles got captured then, their energy would {\em eternally} dominate over baryonic matter, leaving no room other DM scenarios. And if the capture occurs at galactic or smaller scale, then the difficulty concerning structure formation would be avoided. 

The most obvious example of capture-able massless particles is the gluons, which, due to their non-linear interaction with quarks and with themselves, cannot even travel freely from one nucleon to another. Certainly the gluonic energy is counted into hadronic matter, and it is not the DM we are looking for. The most probable massless particles which may be bounded at cosmological scale are the gravitons, which, very much like the gluons, interact non-linearly with matter and with themselves.

The actual history of the universe favors qualitatively the above DM scenario of captured massless particles. At the beginning stage of structure formation, the energy densities of matter and radiation were comparable. 
If the capture happened around this period, then today we would indeed find the total energy of DM comparable to, but higher than that of baryonic matter, because massless particles undergo blue-shift when galaxies were formed through contraction of dilute gases. 

This work was support by China NSF grant 10475057.

\end{document}